\documentclass[aps,prl,twocolumn,superscriptaddress, showpacs]{revtex4}

\usepackage[ansinew]{inputenc}
\usepackage{graphicx}
\usepackage{amssymb}

\newcommand{\chemie}[1]{\ensuremath{\mathrm{#1}}}
\newcommand{\eqdef}{\stackrel{\scriptscriptstyle\wedge}{=}}

\begin{document}


\title{Transport through (Ga,Mn)As nanoislands: Coulomb-blockade and temperature dependence of the conductance}


\author{Markus Schlapps}
\affiliation{Institut f\"{u}r Experimentelle und Angewandte Physik,
Universit\"{a}t Regensburg, 93040 Regensburg, Germany}
\author{Teresa Lermer}
\affiliation{Institut f\"{u}r Experimentelle und Angewandte Physik,
Universit\"{a}t Regensburg, 93040 Regensburg, Germany}
\author{Stefan Geissler}
\affiliation{Institut f\"{u}r Experimentelle und Angewandte Physik,
Universit\"{a}t Regensburg, 93040 Regensburg, Germany}
\author{Daniel Neumaier}
\affiliation{Institut f\"{u}r Experimentelle und Angewandte Physik,
Universit\"{a}t Regensburg, 93040 Regensburg, Germany}
\author{Janusz Sadowski}
\affiliation{Max-Lab, Lund University, SE-223 63 Lund, Sweden}
\affiliation{Institute of Physics, Polish Academy of Sciences, 02-668 Warsaw, Poland}
\author{Dieter Schuh}
\affiliation{Institut f\"{u}r Experimentelle und Angewandte Physik,
Universit\"{a}t Regensburg, 93040 Regensburg, Germany}
\author{Werner Wegscheider}
\affiliation{Institut f\"{u}r Experimentelle und Angewandte Physik,
Universit\"{a}t Regensburg, 93040 Regensburg, Germany}
\affiliation{Laboratory for Solid State Physics, 8093 Zurich, Switzerland}
\author{Dieter Weiss}
\affiliation{Institut f\"{u}r Experimentelle und Angewandte Physik,
Universit\"{a}t Regensburg, 93040 Regensburg, Germany}

\date{\today}

\begin{abstract}
We report on magnetotransport measurements of nanoconstricted (Ga,Mn)As devices showing very large resistance changes that can be controlled by both an electric and a magnetic field. Based on the bias voltage and temperature dependent measurements down to the millikelvin range we compare the models currently used to describe transport through (Ga,Mn)As nanoconstrictions. We provide an explanation for the observed spin-valve like behavior during a magnetic field sweep by means of the magnetization configurations in the device. Furthermore, we prove that Coulomb-blockade plays a decisive role for the transport mechanism and show that modeling the constriction as a granular metal describes the temperature and bias dependence of the conductance correctly and allows to estimate the number of participating islands located in the constriction.
\end{abstract}


\maketitle

Narrow constrictions in thin stripes of the ferromagnetic semiconductor (Ga,Mn)As display huge magnetoresistance (MR) effects \cite{ruester2003:TMR, giddings2005:TAMR, schlapps2006, ciorga2007:TAMR, pappert2007:NVMD, wunderlich2006:CBAMR}, applicable, e.g., for sensors or non-volatile memory elements \cite{pappert2007:NVMD}. The underlying mechanism causing these MR effects is still not unambiguously resolved. Initially it was believed that collinear alignment of the magnetization on both sides of the constriction together with a tunnel barrier, formed by side wall depletion in the narrow, gives rise to a tunneling magnetoresistance effect (TMR) \cite{ruester2003:TMR}. In these experiments the in-plane magnetic field was aligned along the axis of the stripes. After the finding that the resistance as a function of direction and strength of an in-plane magnetic field is reminiscent of the anisotropic magnetoresistance (AMR), the observed MR effects were ascribed to a tunneling anisotropic magnetoresistance effect (TAMR) \cite{giddings2005:TAMR}, originally observed in an Au/\chemie{AlO_x}/(Ga,Mn)As sandwich structure \cite{gould2004:TAMR}. There, the angular dependence arises from spin-orbit coupling which results in a magnetization dependent density of states (DOS) \cite{giddings2005:TAMR}. However, in the case of lateral structures, the variation in the DOS is $\sim$50\% \cite{giddings2005:TAMR} and much too small to explain the MR effects, orders of magnitude larger \cite{schlapps2006, ciorga2007:TAMR}. A possible explanation was based on the assumption that a metal insulator transition (MIT) occurs \cite{pappert2006:MIT, pappert2007:NVMD}, again driven by spin-orbit interaction and a magnetization dependent overlap of hole wavefunctions. On the other hand, experiments on a narrow (Ga,Mn)As channel revealed a Coulomb-blockade anisotropic magnetoresistance effect (CBAMR), where the angular dependence of the resistance is ascribed to chemical potential anisotropies \cite{wunderlich2006:CBAMR}. Here we revisit the problem of transport across a (Ga,Mn)As nanoconstriction. In order to shed light on the underlying transport mechanism, the bias-, temperature- and gate-dependence of the conductance is in the focus of the present letter. After a phenomenological explanation for the large MR effects and the evidence of Coulomb-blockade (CB) we regard the constricted region as a granular metal and resort to a theoretical model to fit the measured bias- and temperature dependence of the conductance.
\begin{figure}
\includegraphics{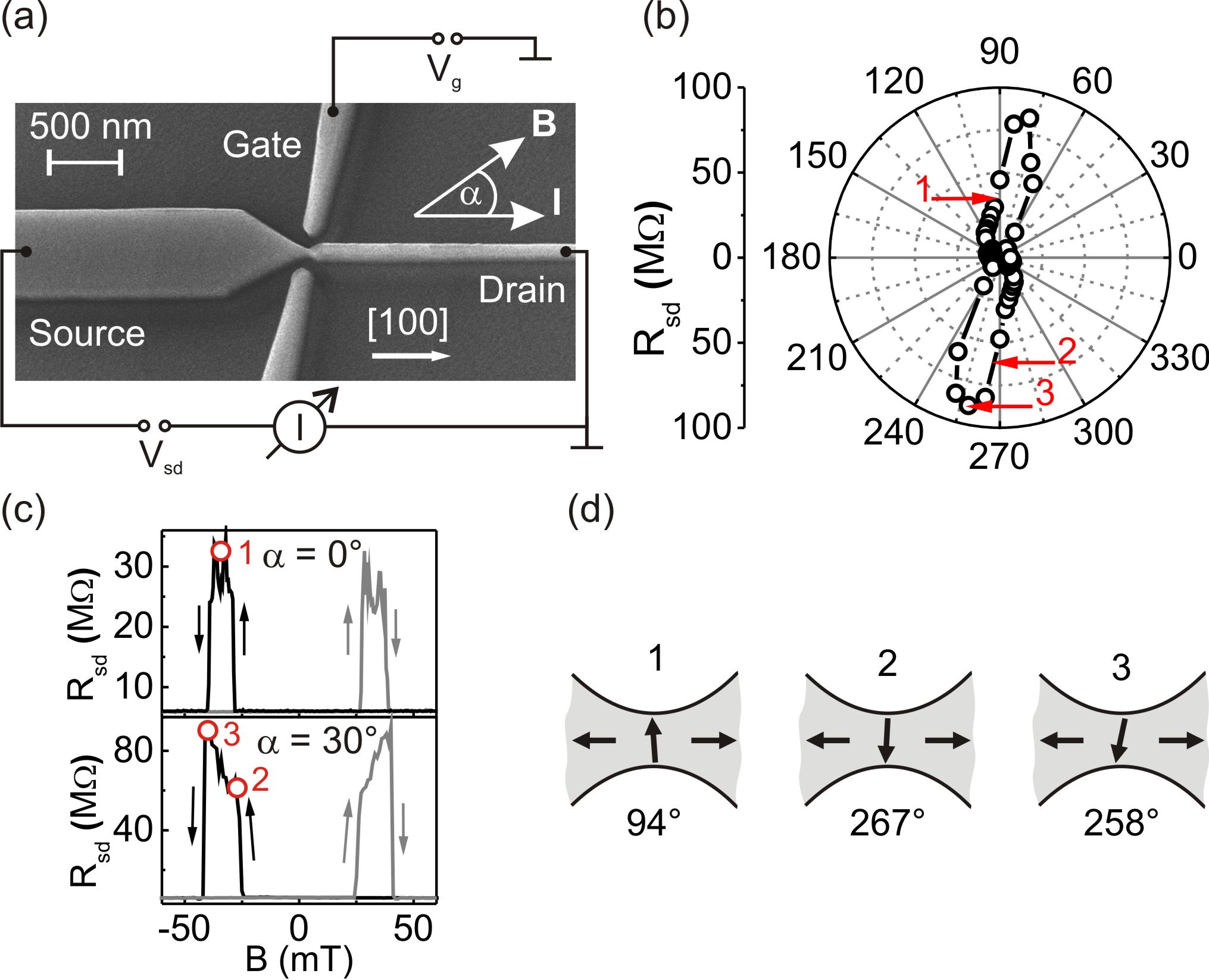}
\caption{\label{figure1}(color online) (a) Electron micrograph of the central part of a device, tilted by 40°. (b) Polar plot of $R_{sd}$ at \mbox{1.6 K} showing the strong anisotropy of $R_{sd}$ for sample A as a function of the magnetization direction. The measurement was done in a high magnetic field of \mbox{300 mT}. (c) MR of sample A for $\alpha = 0°$ and $\alpha = 30°$ at \mbox{1.6 K}. The switching fields of $\pm\mbox{28 mT}$ and $\pm\mbox{39 mT}$ (for $\alpha = 0°$) correspond to the magnetization reversal of the broad and narrow lead, respectively Comparing the resistance values of (b) and (c) allows to deduce the magnetization alignment in the constriction. The configurations for a HR state together with the magnetization angle in the constriction is sketched in (d).}
\end{figure}

We fabricated samples with an individual nanoconstriction; an electron micrograph of the central region of one of the devices is shown in Fig.~\ref{figure1}(a). The device consists of a 20 nm thick \chemie{Ga_{0.95}Mn_{0.05}As} film grown at 243°C by low temperature molecular beam epitaxy on top of \mbox{5 nm} \chemie{Al_{0.7}Ga_{0.3}As}, \mbox{3 nm} LT GaAs and a (001)-GaAs substrate. After annealing the unpatterned sample at 200°C for \mbox{8.5 h} the Curie temperature, determined from the temperature dependence of the resistance \cite{Novak2008:curie}, reached approximately \mbox{90 K} at a carrier density of about $1.8 \times 10^{20} cm^{-3}$. The central area including the gates were defined by electron beam lithography where cross-linked poly-methyl-methacrylate (PMMA) formed the etch mask. Chemically assisted ion beam etching was then used to define the (Ga,Mn)As structure. Ti/Au contacts (not shown in Fig.~\ref{figure1}(a)) were deposited as source-drain and gate electrodes. Each of the devices consists of a \mbox{3 $\mu$m} long and \mbox{700 nm} wide lead separated by a \mbox{$\sim$20 nm} wide constriction from a \mbox{3 $\mu$m} long and \mbox{100 nm} wide lead. The structure is aligned along the [100]-direction, that is close to an easy axis of (Ga,Mn)As films. Apart from that, strain relaxation in the stripes turns the easy axis towards the leads \cite{wenisch2007:anisotropy}. To determine the switching fields for a magnetization reversal in the wide and narrow stripes via the AMR effect some of the devices have additional potential probes on each stripe. Below we present data of two samples A and B. While sample A has no side gate, sample B has additional side electrodes (as presented in Fig.~\ref{figure1}(a)), separated by \mbox{$\sim$130 nm} from the constriction, to tune the electrostatic potential in the constriction. Magnetotransport measurements were carried out in a \textsuperscript{4}He bath cryostat or a \textsuperscript{3}He/\textsuperscript{4}He dilution refrigerator, each equipped with a superconducting magnet. The angle $\alpha$ between the applied in-plane magnetic field and the current direction could be varied by rotating the sample inside the cryostat. The measurements were done using dc-technique in a two-point configuration: a constant voltage $V_{sd}$ between source and drain was applied and the resulting current $I$ was measured employing a current amplifier.

First we demonstrate that the MR effects observed in (Ga,Mn)As nanoconstrictions can be phenomenologically explained by the magnetization alignment in the constriction. Figure \ref{figure1}(b) shows a polar plot of the resistance $R_{sd}$ of sample A measured at \mbox{1.6 K} by rotating the sample in an in-plane magnetic field of \mbox{300 mT}. This field strength is large enough to align all magnetization vectors into the external field direction. $R_{sd}$ is strongly anisotropic with resistance changes of up to a factor of 20. Neither AMR, which in bulk (Ga,Mn)As is on the order of a few percent, nor TMR (magnetization in the leads is always parallel) or TAMR (change of DOS too small) can explain such drastic resistance changes \cite{ciorga2007:TAMR}. Figure \ref{figure1}(c) displays the MR as a function of the in-plane field during a magnetic field sweep along the [100]-structure axes ($\alpha = 0°$) and for the direction causing the largest MR effect ($\alpha = 30°$). The switching fields of the spin-valve like signal perfectly agree with the magnetization reversal fields of the \mbox{700 nm} and \mbox{100 nm} wide stripe that have been detected simultaneously by a four-point measurement using a sample with additional voltage probes on each lead (not shown here). Note that the switching fields are different in the wide and narrow stripe due to a different strain relaxation. The large resistance change in Fig.~\ref{figure1}(c) can therefore be ascribed to the consecutive magnetization reversal of the \mbox{700 nm} and the \mbox{100 nm} stripe. The maximum resistance observed in the high field experiment of Fig.~\ref{figure1}(b) is in good agreement with the high resistance (HR) measured in the experiment of Fig.~\ref{figure1}(c). This suggests that both effects stem from the same origin and that $R_{sd}$ is linked to the magnetization direction in the constriction. Thus, by comparing the resistances of the polar plot of Fig.~\ref{figure1}(b)with the spin-valve like signal observed during a magnetic field sweep we can deduce the magnetization direction in the constriction and explain the MR trace by means of the magnetization alignments in the device. The magnetization configurations for the HR states marked with open circles in Fig.~\ref{figure1}(c) are sketched in Fig.~\ref{figure1}(d). So, at low external magnetic fields, the relative alignment of the magnetization in the wide and narrow stripe involves a distinct magnetization orientation in the constriction and thus determines the resistance. Within this picture the MR is easily explainable for an applied magnetic field along the stripe axis (i.e., $\alpha = 0°$). Here, the HR state occurs due to the 180° magnetization reversal in the \mbox{700 nm} stripe at \mbox{-28 mT} whereas the magnetization in the \mbox{100 nm} stripe still remains in the [100] easy axis [see sketch 1 in Fig.~\ref{figure1}(d)] until its coercitive field is reached at \mbox{-39 mT}. The antiparallel alignment of the stripes causes the HR due to the associated orientation of the magnetization in the constriction and not due to the TMR effect. But also more complex features of the MR trace that appear for magnetic field angles $\alpha \neq 0°$ can be understood. Consider, e.g., the MR in the case of $\alpha = 30°$, shown in Fig.~\ref{figure1}(c). The resistance increase within the magnetic field range marked with 2 and 3 arises from a coherent magnetization rotation in the constriction due to an increasing magnetic field strength along the 210° ($\eqdef -30°$) direction. This is illustrated in the cartoons 2 and 3 of Fig.~\ref{figure1}(d). The magnetizations in the leads remain unchanged in the considered magnetic field window due to the strong uniaxial anisotropy of the stripes. Increasing the magnetic field strength further leads to an abrupt 180° magnetization reversal in the \mbox{100 nm} stripe causing the abrupt resistance change due to the resulting magnetization alignment in the constriction towards 180°.

We now address the microscopic origin of the large MR effects and discuss the results with respect to the available models, magnetization induced MIT as proposed in \cite{pappert2007:NVMD} and Coulomb-blockade assisted as reported in \cite{wunderlich2006:CBAMR}. At temperatures below \mbox{$T \sim 30$ K} the resistance of our patterned samples increases continuously with decreasing temperature (not shown here), both for the high resistance (HR) and low resistance (LR) state, thus indicating that the huge MR effects, which vanish above \mbox{$\sim30$ K}, only exist in the insulating regime. A picture involving a magnetization driven MIT, for which a different temperature behavior for the HR and LR state would be expected, is therefore at odds with our experiment. In the CBAMR picture \cite{wunderlich2006:CBAMR}, however, the system is expected to be insulating both in the HR and LR state and is consistent with our results. The model is based on the assumption that disorder potential fluctuations together with side wall carrier depletion create small, isolated islands in the constriction.

\begin{figure}
\includegraphics{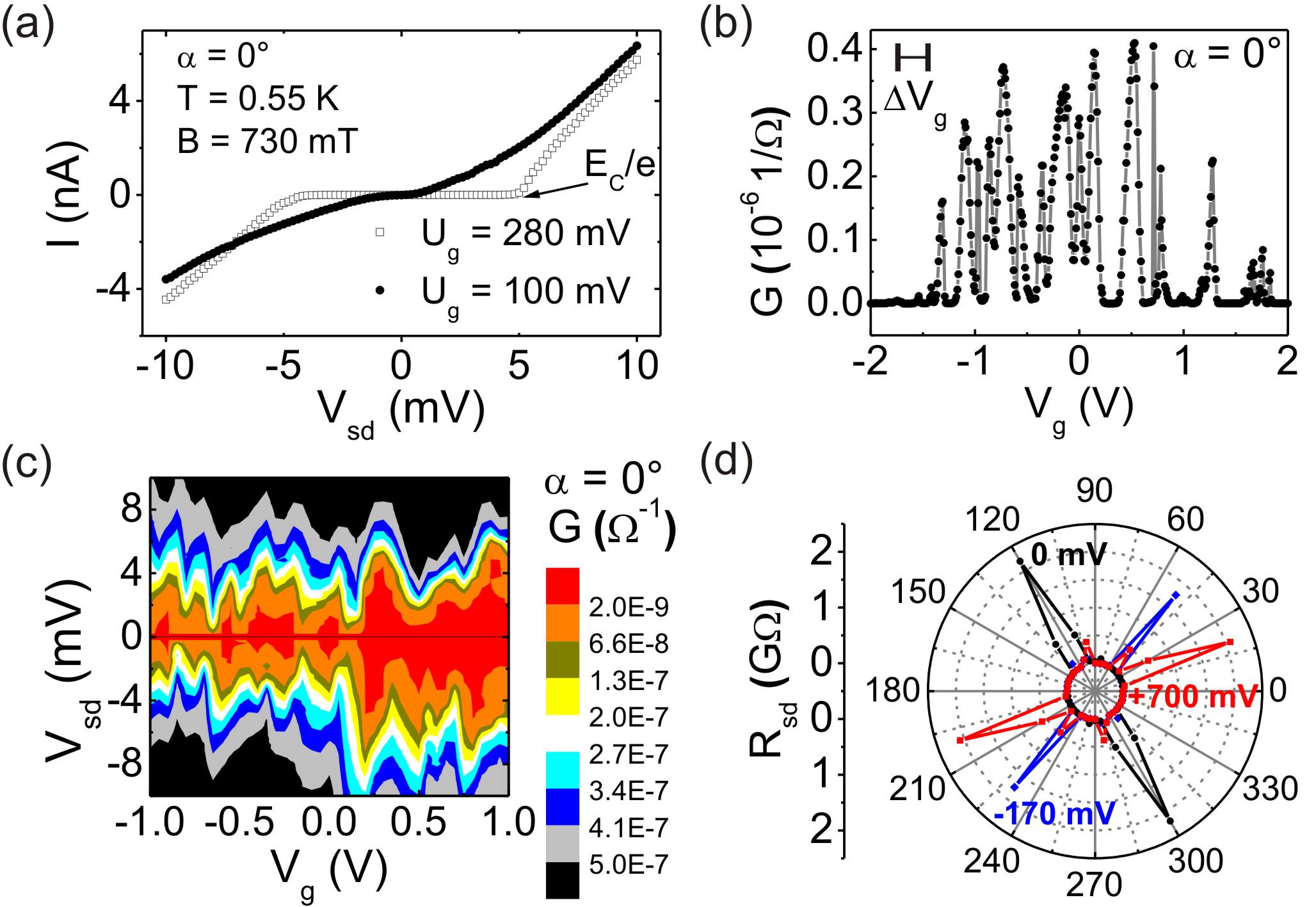}
\caption{\label{figure2}(color online) (a) $I$-$V$ characteristics of sample B for two different gate voltages showing that the device acts as a transistor. (b) Dependence of the conductance on the gate voltage for \mbox{$\alpha$ = 0°} and \mbox{$V_{sd}$ = 3 mV}. The oscillating behavior is ascribed to a CB transport. $\Delta V_g$ stands for the average oscillation period. (c) Color plot of the conductance versus gate and source-drain voltage resulting in a CB diamond-like structure. The diamond pattern changes for different magnetization directions, here shown for 0°. (d) Polar plot of $R_{sd}$ for different gate voltages. All measurements were carried out at $B$ = \mbox{730 mT} and $T$ = \mbox{550 mK}.}
\end{figure}

To prove that in our (Ga,Mn)As point contacts Coulomb-blockade is at work we investigated samples with additional side-gate electrodes. The non-linearity of the $I$-$V$ characteristic of device B, plotted in Fig.~\ref{figure2}(a), clearly depends on the gate voltage. The corresponding conductance $G$ [Fig.~\ref{figure2}(b)], measured in an in-plane field of \mbox{0.73 T} \footnote{The $B$-field that forces the magnetizations into the field direction is larger for lower temperatures.} along the [100]-direction at \mbox{$T = 0.55$ K}, displays pronounced oscillations, indicating Coulomb-blockade. These conductance oscillations have an irregular spacing, suggesting that more than one island is formed in the constriction. From the average oscillation period $\Delta V_g$ in Fig.~\ref{figure2}(b) we estimate a capacitance \mbox{$C_g \sim$ \mbox{0.55 aF}} between gate electrode and islands. For a very rough estimate of the island area we use the charging energy of some meV taken from the $I$-$V$-characteristics. This gives a total capacitance $C_\Sigma$ of order \mbox{10 aF}. By approximating the island as a sphere we thus expect a radius of order \mbox{10 nm} consistent with the constriction size of our device. The pronounced CB diamond-like structure in Fig.~\ref{figure2} (c) highlights the role of CB in transport across (Ga,Mn)As point contacts. The anisotropy of the resistance and its dependence on $V_g$ is most clearly seen in Fig.~\ref{figure2}(d) where polar plots of $R_{sd}$ are shown for three gate voltages. With different gate voltage the direction of $B$ and hence of the magnetization where the largest resistance is observed changes. Hence, both the gate voltage and the magnetization are independent parameters which allow us to tune the conductance through ferromagnetic (Ga,Mn)As islands. The underlying physics has been described by Wunderlich et al.~\cite{wunderlich2006:CBAMR} within the CBAMR model and is due to (different) changes in the chemical potential inside and outside of conducting (Ga,Mn)As islands. Due to the strong spin-orbit coupling in (Ga,Mn)As the chemical potential in the island depends on both the gate voltage and the magnetization direction. These effects might be enhanced by the magneto-Coulomb effect \cite{Ono1997:enhanced}. The different resistance anisotropy for devices A and B at \mbox{$V_g$ = 0} (see Fig.~\ref{figure1}(b) and \ref{figure2}(d)) can be explained by a different electrostatic environment in the vicinity of the constriction due to thermal cycling \cite{giddings2008:TAMR2}.

\begin{figure}
\includegraphics{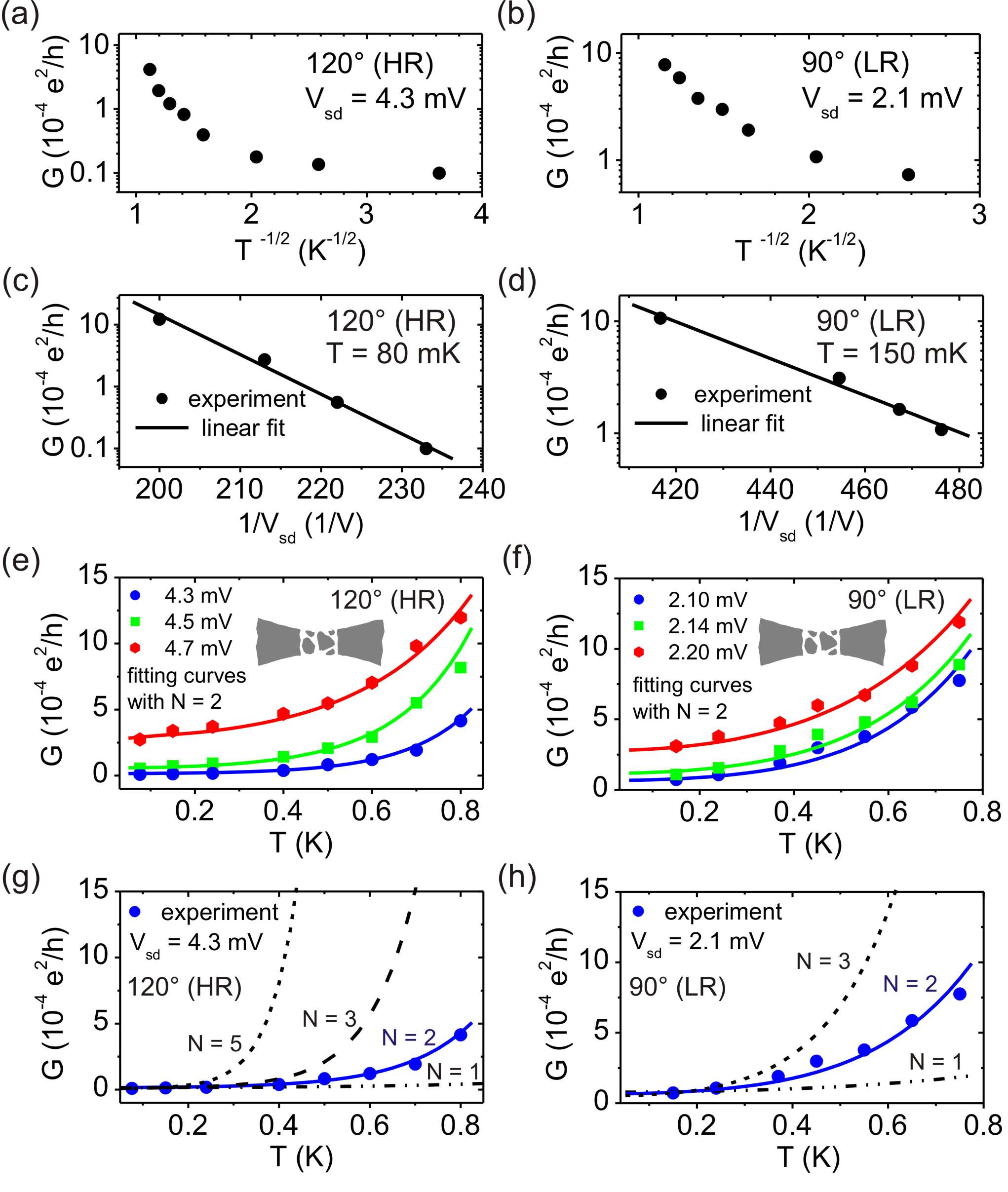}
\caption{\label{figure3}(color online) [(a - b)] $\log G$ versus $T^{-1/2}$ plot in a HR and LR state realized by applying a $B$-field of \mbox{730 mT} along 120° and 90°, respectively, showing that our experiments are not consistent with the low-field regime of the model for granular metal films. [(c - d)] Voltage dependence of $G$ showing a good agreement with the model for the high-field regime. From the linear fit we obtain $\ln \left[G_\infty (HR)/10^{-4}e^2/h\right] = 32 \pm 2$, $V_0 = (0.15 \pm 0.01)$V for $\alpha$ = 120° (HR) and $\ln \left[G_\infty (LR)/10^{-4}e^2/h\right] = 18 \pm 2$, $V_0 = (0.038 \pm 0.003)$V for $\alpha$ = 90°. [(e - f)] Experimental data and numerical fits for the temperature dependence of $G$ for different $V_{sd}$ for HR and LR state, respectively. $N$ is the fitting parameter and indicates the number of islands which are schematically illustrated in the inset. The sensitivity of $N$ is demonstrated in (g) and (h).}
\end{figure}

As a result of the gate measurements we conclude that more than one conducting island is located within the nanoconstriction. To elucidate the transport mechanism further we studied the temperature and bias voltage dependence of the conductance $G$ across the constriction. Corresponding data are shown in Fig.~\ref{figure3} for both, HR and LR state. While in Figs. \ref{figure3} (a) and (b) the conductance is plotted as a function of $1/\sqrt{T}$, Figs. \ref{figure3}(e) and (f) display $G$ as function of $T$ for different source-drain voltages $V_{sd}$. The conductance shows a strong bias dependence [see also Figs.~\ref{figure3}(c) and \ref{figure3}(d)]. Such behavior is expected once Coulomb-blockade is involved in the transport process. As our system is reminiscent of a granular metal film - metal grains separated by insulating shells - which have been investigated intensively in the past (see, e.g., \cite{abeles1975:granular} and references therein) - we resorted to a corresponding model. We went back to the model of Abeles et al.~\cite{abeles1975:granular} and applied it to our system. Due to the surface depletion in the nanoconstriction we assume that metallic spots of (Ga,Mn)As are separated from each other by thin insulating regions, thus justifying the granular metal assumption. As shown below the modeled bias and temperature dependence of the conductance fits perfectly the experimentally observed data. The granular model distinguishes two regimes reflecting the interplay of the involved energy scales: the thermal energy $k_BT$, the charging energy \mbox{$E_C = e^2/2C_\Sigma$} and the energy associated with the average voltage drop across one island $e\Delta V = eV_{sd}/N$. Here, $N$ indicates the number of islands whereat the voltage drop occurs. For low electric fields where $e\Delta V$ is much smaller than $k_BT$ the model is mainly based on charge-carrier generation by thermal activation and predicts: $G(T) = G_0 \exp\{-2(C/\sqrt{k_BT})\}$. Here, $C$ is a constant proportional to the charging energy. However, our experiments are clearly not in this low-field regime since the data do not follow the $\ln G \varpropto T^{-1/2}$ dependence over the entire temperature range shown in Figs.~\ref{figure3}(a) and \ref{figure3}(b), both for the HR and LR state. This is consistent with a simple estimate: for our highest temperature of \mbox{0.8 K} corresponding to $k_BT/e \sim \mbox{0.07 mV}$, and lowest applied bias voltage, $V_{sd} = \mbox{2.1 mV}$, the low field regime requires that the number of islands $N$ is much larger than 30. Considering the dimensions of our constriction this is highly unlikely. Thus, we have to examine the high-field regime ($e\Delta V \geq k_BT$) where field induced tunneling becomes important and increases the carrier density. In this case the conductance is given by \cite{abeles1975:granular}
\begin{eqnarray} \label{high_field_T} G(T, V_{sd})&=&G_\infty \exp(-\frac{V_0}{V_{sd}})\nonumber\\*
 & &\times
\int_{-\frac{V_0}{V_{sd}}}^\infty dZ\frac{Z \cdot
\exp(-Z)}{1-exp\left\{-\frac{Z \cdot e \cdot
(V_{sd}/N)}{(Z+\frac{V_0}{V_{sd}})k_BT}\right\}}\hspace{0.5cm}
\end{eqnarray}
with the electron charge $e$ and the constant $V_0$ depending on the charging energy. In the limiting case of zero temperature the conductance is given by
\begin{equation} \label{high_field_V}
G(V_{sd}) = G_\infty \exp\left(-V_0/V_{sd}\right)
\end{equation}
Hence, plotting $\ln G$ versus $1/V_{sd}$ at the lowest temperature allows to estimate both $G_\infty$ and $V_0$ directly from experiment. The corresponding data are shown for the HR and LR state in Fig.~\ref{figure3}(c) and (d), respectively. The only undetermined parameter in Eq.~\ref{high_field_T} is the number of islands $N$ which we use as the only free fit parameter. The experimental data at temperatures below \mbox{1 K}, plotted for different bias voltages and for HR and LR states in Figs.~\ref{figure3}(e) and \ref{figure3}(f), respectively, can be very well fitted by using $N = 2$. Eq.~\ref{high_field_T} describes the temperature dependence for all three traces correctly, both for HR and LR state, using within the error margins the same values for $G_\infty$ and $V_0$ given in the caption of Fig.~\ref{figure3}. The high sensitivity of the fits with respect to the number of islands is demonstrated in Figs.~\ref{figure3}(g) and (h). Assuming that the constriction contains two islands and using the previously estimated island size of \mbox{20 nm} diameter suggests an individual island diameter of order \mbox{10 nm}. The granular metal model, though at the fringe of applicability given the low $N$ value, describes surprisingly well the bias and temperature dependence of the conductance. This provides further evidence that Coulomb-blockade plays a prominent role in transport through (Ga,Mn)As nanoconstrictions and is responsible for the huge magnetoresistance effects observed.
\\

\begin{acknowledgments}
We thank Rashid Gareev for valuable discussions and acknowledge support from the Deutsche Forschungsgemeinschaft (DFG) via SFB 689.
\end{acknowledgments}


\end{document}